# MODELING ENTERPRISE ARCHITECTURE USING TIMED COLORED PETRI NET: SINGLE PROCESSOR SCHEDULING

Saied Pashazadeh[1] and Elham Abdolrahimi Niyari[2]

[1]Faculty of Electrical and Computer Engineering, University of Tabriz, Tabriz, Iran

[2]Department of Management IT, University College of Mizan, Tabriz, Iran

## ABSTRACT

*The purpose of modeling enterprise architecture and analysis of it is to ease decision making about architecture of information systems. Planning is one of the most important tasks in an organization and has a major role in increasing the productivity of it. Scope of this paper is scheduling processes in the enterprise architecture. Scheduling is decision making on execution start time of processes that are used in manufacturing and service systems. Different methods and tools have been proposed for modeling enterprise architecture. Colored Petri net is extension of traditional Petri net that its modeling capability has grown dramatically. A developed model with Colored Petri net is suitable for verification of operational aspects and performance evaluation of information systems. With having ability of hierarchical modeling, colored Petri nets permits that using predesigned modules for smaller parts of the system and with a general algorithm, any kind of enterprise architecture can be modeled. A two level hierarchical model is presented as a building block for modeling architecture of Transaction Processing Systems (TPS) in this paper. This model schedules and runs processes based on a predetermined non-preemptive scheduling method. The model can be used for scheduling of processes with four non-preemptive methods named, priority based (PR), shortest job first (SJF), first come first served (FCFS) and highest response ratio next (HRRN). The presented model is designed such can be used as one of the main components in modeling any type of enterprise architecture. Most enterprise architectures can be modeled by putting together appropriate number of these modules and proper composition of them.*

## 1. INTRODUCTION

Enterprise Architecture (EA) is a set of models that are created based on the descriptions of an organization. It is created to match the needs of management and better maintenance of system in lifetime of a system. John Zachman is one of the pioneers in architecture of information systems and EA. He believes that without appropriate architecture we cannot effectively use information technologies [1]. Timing and scheduling of processes is one of the rows of Zachman's architecture. In information systems like Transaction Processing Systems (TPS), scheduling is important factor that greatly influences the efficiency of a system. Zachman framework for enterprise architecture can be used as a guideline for modeling architecture of any information system. Its framework can be used for determining scope and amount of details that each aspect of enterprise architecture must be modeled.

                                                1

Main aim of this paper is providing a basis for modeling any type of enterprise architectures. Achieving this idea requires design of few basic parameterized and general building blocks. This paper only deals with design and implementation of one of these basic blocks. In this paper, scheduling and running of processes by a single processor with four non-preemptive scheduling methods are modeled. Scheduling of a single processor as basic building blocks in modeling of complex enterprise architectures are under study in this paper. Colored Petri net is selected in this paper from wide range of modeling techniques and tools that are used for modeling enterprise architectures.

Colored Petri net is an extension of classical Petri net that tokens must have color type [2]. In presented model of scheduling a single processor in this paper, complex color sets like arrays of records are applied. Using complex color sets simplifies the model and increases its performance in comparison with similar models that do not used complex colors. Presented model accepts processes, their input times and service time beyond name of a single non-preemptive scheduling method as input and automatically executes the processes. This model computed waiting time, turnaround time of processes and idle times of the processor. Model is designed so that all the necessary details regarding scheduling and running processes can be obtained for subsequent processing, and goal of the model is not only running of processes.

## 2. RELATED WORK

Computer-based information systems are used in four fields of 1) input data, 2) data storage, 3) data processing and 4) information output of any systems. Computer processing is based a collection of programs that support major operations of the system. During the last decade, using enterprise architecture has grown and established for management of information systems in an organization. Enterprise architecture is model-based [3] and its modelling is very complex. Macro architecture layer of EA includes mission, goals and strategic plans. Operation layer includes workflow, information flow and financial flow. Planning and scheduling is very important in the operational layer [4].

Main purpose of a workflow management system is supporting of definition, execution, registration and control of processes. Because processes are a dominant factor in workflow management, it is important to use an established framework for modelling and analysing workflow processes. Carl Adam Petri invented the classical Petri net in the sixties. Since then Petri nets have used to model and analyse wide range of systems with applications ranging from protocols, hardware, and embedded systems to flexible manufacturing systems, user interaction, and business processes. In the last two decades, the classical Petri net extended with colour, time and hierarchy. These extensions facilitate the modelling of complex systems where data and time are important factors in them [5]. There are three good reasons for using Petri nets for workflow modelling and analysis: (1) formal semantics despite the graphical nature, (2) being state-based instead of event-based and (3) an abundance of analysis techniques and tools [6]. Petri nets are widely used for modeling time-dependent processes and distributed flow systems such as data communications networks and manufacturing processes [7]. For modeling of workflow, modeler can use ready patterns such as pattern of parallel processes, sequence of serial processes, and single-process workflow and many similar patterns to convert them into Petri nets [8]. Petri net models of non-preemptive and preemptive schedulers with additions of resource reservation and aperiodic service are presented in [9]. The scheduling problems in general can be modeled using continuous or discrete time models and is typically driven by feasibility and focuses on short term time horizon [10].

There are many types of system but essentially all systems are made up of various parts that are connected together in a particular ways such that they can interact appropriately with each other to achieve specific purposes. Way of representing different elements and behaviors of production systems with the use of the HTCPN formalism and how to build complete models of production system is proposed in [11]. It is possible to point two main kinds of application of the HTCPN formalism in production modelling, simulation and scheduling. (1) Models can be built and simulated using CPN Tools. The design process of a net model is fast and convenient in this case. This approach can be used for finding and verification of production system properties. (2) The HTCPN formalism can be used for creating a new algorithm of simulation and scheduling. Using HCPN eases modelling of large-scale complex systems, and establish a good foundation for modular and hierarchical design for future development of system [12]. We proposed a Petri net based method to give a comprehensive model which can represent the complex constraints like batches, setup times, transportation and multiple resources in real-life job shop scheduling problems. The general occurrence graph and simulation analysis method are introduced to find the best solution based on CPN tools [13]. Time and scheduling is an aspect of Zachman framework in enterprise architecture framework [14].

In this paper a hierarchical timed coloured Petri nets (HTCPN) formalism of single processor scheduling with nonpreemtive scheduling methods names FCFS, SJF, HRRN and PR is presented. Model of single machine is important for various reasons. The single machine environment is very simple and is a special case of all other environments. Model of single machine can be used for modelling parallel, series and various other combinations of processes in a system. The results that can be obtained for single machine models not only provide insights into the single machine environment, they also provide a basis for heuristics that are applicable to more complicated machine environments [15]. Genetic algorithm approach can be used for finding near optimal scheduling of machines [16].

## 3. COLOUR SETS, INITIAL MARKINGS AND MODEL OF SYSTEM

In this part of paper, brief descriptions of colour sets, variables, initial markings and colored Petri net model of system are presented.

### 3.1. Color Sets

Color sets that used in modeling of single processor scheduling are as follows:
colset RT= product INT*INT;
colset Process = record PI:INT * IT:INT * ST:INT * WT:INT * ES: INT * PR:RT timed;
colset ProcList = list Process timed;
colset SCHTYPE = with FCFS | SJF | PR | HRRN timed

The colour set *RT* is defined as two pairs of integer numbers that represents priority of a process. First field is considered as major priority and second field as minor priority. Some scheduling methods produce real values as priority of a process. Most of the colored Petri net tools do not supports real numbers. Therefore, priority of a process is defined as two integer numbers. First number represents integer part of process's priority and second integer number represents fractional part of process's priority. Color set *Process* is a record that contains six fields. First field denoted with *PI* is of type integer and represents process index. Second field is denoted with the title *IT* is of type integer and represents arrival time of a process to system for running. Third field is denoted with title *ST* and its type is integer and represents service time of the process. Fourth field is denoted with title *WT* and is of type integer and represents waiting time of a process until current simulation time. Fifth field is denoted with the title *ES* is of type integer and represents execution start time of the process. Last field is denoted with *PR* and is of type *RT*. This field represents current priority of the process. Color set *ProcList* defines a list of elements of type *Process*. Colour set *SCHTYPE* is used to represent types of non-preemptive scheduling that can be used in the model. It is enumerated type of FCFS (First Come First Served), SJF (Shortest Job First), PR (Priority Based) and HRRN (Highest Response Ratio Next). Timed model of system is developed and therefore all types have been defined as timed.

## 3.2. Initial Marking and Variables

Initial marking of the model is defined as follows:
val InitialProcess =
[{PI=1, IT=6, ST=4, WT=0, ES=0, PR=(2,0)},
 {PI=2, IT=7, ST=3, WT=0, ES=0, PR=(1,0)},
 {PI=3, IT=8, ST=2, WT=0, ES=0, PR=(2,0)},
 {PI=4, IT=5, ST=2, WT=0, ES=0, PR=(3,0)},
 {PI=5, IT=9, ST=3, WT=0, ES=0, PR=(1,0)},
 {PI=6, IT=1, ST=3, WT=0, ES=0, PR=(4,0)}]

This initial marking represents six processes that are used as case study in this paper. Table 1 show index, input times, required service time and static predetermined priority of processes.

Table 1.  Input processes of a processor

| Process Index | Input Time | Service Time | Priority |
|---------------|------------|--------------|----------|
| P1 | 6 | 4 | 2 |
| P2 | 7 | 3 | 1 |
| P3 | 8 | 2 | 2 |
| P4 | 5 | 2 | 3 |
| P5 | 9 | 3 | 1 |
| P6 | 1 | 3 | 4 |

Variables of the model are as follows:
var sm: SCHTYPE;
var process:Process;
var allproc, run, rq, rp, proc, queue: ProcList;

## 3.3. Model of System

Figure 1 shows top level CPN model of single processor's scheduler. Figure 2 show CPN model of single processor with non-preemptive scheduler.

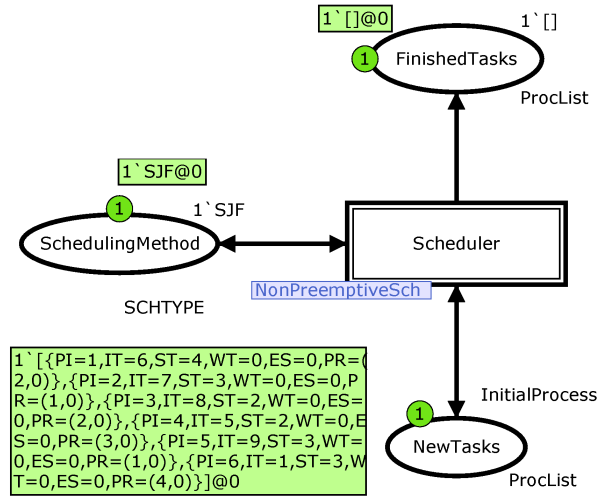

Figure 1. Top level CPN model of single processor's scheduler

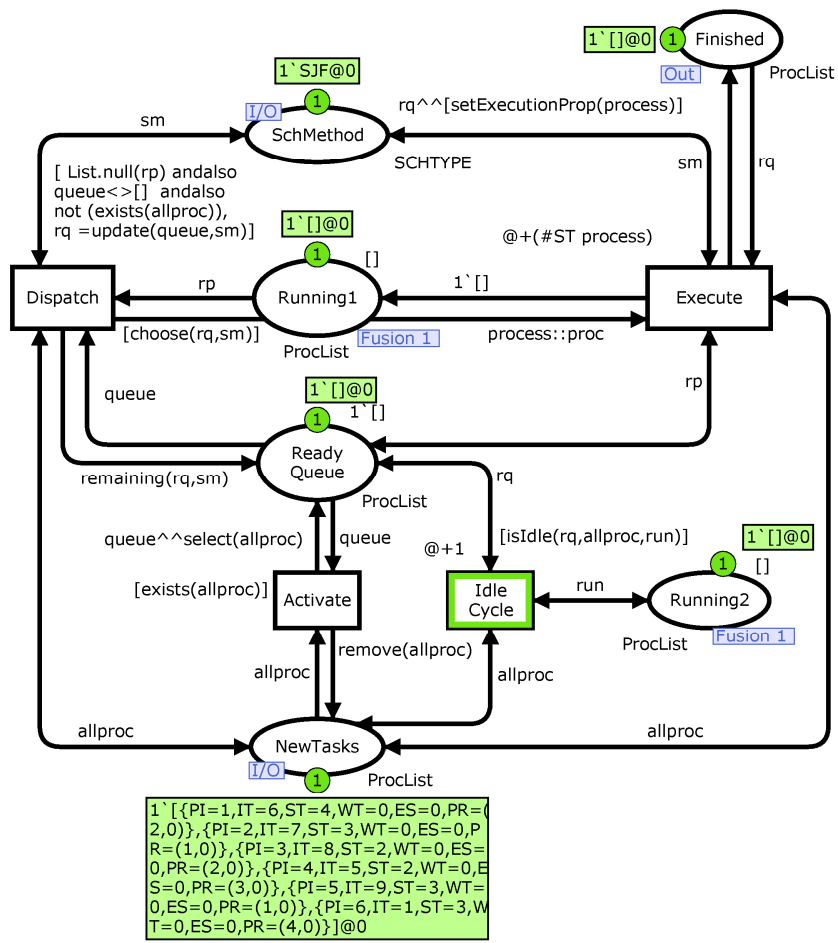

Figure 2. Low level CPN model of single processor with non-preemptive scheduler

# 4. FUNCTIONS OF MODEL

In this part of paper brief description of model's functions are presented. Figure 3 shows the structure chart of model's functions.

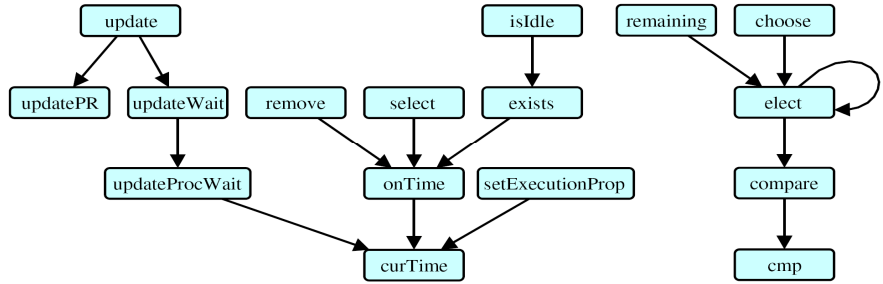

Figure 3. Structure chart of model's functions

Function *curTime* has no input parameter and returns the current simulation time of the model as an integer number.

fun curTime(): int = IntInf.toInt(time());

Function *onTime* takes a process as input parameter and calls function *curTime*. If input time of process is less than or equal to the current simulation time (the input time of process is reached) it returns true as the result and in otherwise returns false.

fun onTime(q:Process): bool = #IT(q)<=curTime();

Function *select* receives the list of processes as input process and produces the list of process as output. This function checks input list of processes from beginning of the list to the end. By calling function *onTime*, if there is a process in the list that its arrival time is reached based on the current simulation time, that process is added to output list of processes.

Function *exists* receives the list of processes as input process and produces a boolean value as result. This function checks input list of processes from beginning of the list to the end. By calling function *onTime*, if there is a process in the list that its arrival time is reached based on the current simulation time, function returns true and in otherwise returns false.

fun exists(z:ProcList) : bool =List.exists onTime z;
Function *remove* receives the list of processes as input process and produces the list of process as output. This function checks input list of processes from beginning of the list to the end. By calling function *onTime*, if there is a process in the list that its arrival time is not reached based on the current simulation time, that process is added to output list of processes.

Function *updatePR* receives two parameters. The first parameter is the scheduling method and second parameter is a process. This function calculates the priority of the inputs process according the scheduling method and updates priority field of that process. In FCFS scheduling method, priority of each process is based on the entry time it to the system. If input time of two processes be equal, then second field of priority will be used. For example, in such case we can give higher priority to the process that is shorter. In SJF method, a process has higher priority that is shorter. If two processes have the same service times, then the one with early input time has higher priority. In PR scheduling methods, priority of processes are defined statically in the beginning. If two processes have the same priority, the one with early input time has higher priority. HRRN priority calculated according to the formula (service time + waiting time) / service time. If two processes have the same major and minor priority values, the one that has lower index number has higher priority.

```
fun updatePR  (sm: SCHTYPE , q:Process) =
   case sm of
     FCFS => Process.set_PR q (#IT q , 0)
   | SJF => Process.set_PR q (#ST q, #IT q)
   | PR => q
   | PR => Process.set_PR q (#1 (#PR q), #IT q)
   | HRRN => Process.set_PR q (((#ST q+ #WT q) * 100) div #ST q,  0)
```

Function *UpdateProcWait* receives a process as input and updates its waiting time field. For scheduling method like HRRN, this field can be used as one of the factors in determining priority of process. This function calls the function *curTime* for receiving current simulation time and decreases the input time of process from it to compute waiting time of that process. This function returns updated input process as output.

```
fun updateProcWait(q:Process)= Process.set_WT q (curTime() - #IT q);
```

Function *updateWait* receive a list of processes and applies function *updateProcWait* on all elements of this list. This function will update waiting times of all the processes of input list.
```
fun updateWait(L:ProcList)=List.map updateProcWait L;
```

Function *update* is one of the important functions that are used in the model. This function receives two parameters. First parameter is a list of the processes that are waiting to be selected for execution. The second parameter determines scheduling method of the model. It updates waiting time of all processes in the input list by calling function *updateWait*. Then it updates priority of all processes in the input list by calling function *updatePR*.
```
fun update(L:ProcList, sm: SCHTYPE) =

let
    val aw=updateWait(L)
    fun  upd L = updatePR( sm,L)
in
  List.map upd aw
end;
```

Function *cmp* takes three parameters as input. The first parameter is calculated priority value for the first process. The second parameter is calculated priority value for the second process. The third parameter represents scheduling method of the model. If scheduling method is FCFS or SJF, Lower numeric value represents a higher priority. If scheduling method is Priority or HRRN, greater numeric value represents higher priority. If the first process has higher priority than second process function returns value of one. If two processes have the same priority, function returns zero value and in otherwise function returns value -1.

```
fun cmp(a: INT, b:INT , sm:SCHTYPE) : INT=
   if ( sm = FCFS orelse sm = SJF ) then
       if  a < b then 1
```

```
        else if a = b then 0 else ~1
else
    if  a > b then 1
    else if a = b then 0 else ~1
```

Function *compare* has three input parameters and return an integer number as result. First and second parameters are the two processes. The third parameter represents scheduling method of the model.  Updated priority of each process is stored in *PR* field of it. This priority value is in the form of a pair of integer values. First numbers is used as the major priority field and the second number is considered as minor field. This function compares major and minor priorities of two input process individually by calling function *cmp*. If major parts from priorities of two processes are equal then, their minor fields are used for determining relative priority of two processes. If both of the major and minor of fields of two input processes are equal, then priority of a process that has lower process index is higher than the other one. If first process has higher priority than the second process, function returns value one and in otherwise function returns -1 as a result.

```
fun compare(p1:Process, p2:Process, sm:SCHTYPE) : INT =
(*if relation > be true for two product then 1
if two product be equal then 0
if relation < be true for two product then ~1 *)
let
    val  r = ref 0
    val  major = cmp( #1 (#PR p1) , #1 (#PR p2) , sm)
    val  minor = cmp( #2 (#PR p1) , #2 (#PR p2) , sm)
in
  if (major <> 0 ) then r:= major
   else  if  (minor <> 0 ) then r:= minor
         else  if (#PI p1 < #PI p2 ) then r:= 1
                  else r:= ~1;
 !r
end
```

Function *elect* is a recursive function with four input parameters. First parameter indicates the list of processes that are ready for running. Second parameter is index of process in the list of first input parameter that has the highest priority till current calling of function *elect*. It should be noted that the index of the processes in the list starts from zero. In first call of the function, the second parameter will be zero and this means that at the beginning, the first process will be considered as the process with highest priority. The third parameter plays the role of the pointer. It represents a process index in the list that in current call will be compared with the process that is specified in the second parameter. This function calls function *compare* based scheduling method on the list of ready processes, select the process that has the highest priority and returns the index it as a result. This function calls the function *compare* and selects the process that has the highest priority index in the list of ready processes based on the scheduling method and returns its index as a result. If multiple processes have the highest priority with same value, then function returns the index of the process that is in head of the process list.

```
fun elect(q:ProcList, maxi:int, c:int, sm:SCHTYPE):INT=
let
    val len = List.length(q)
in
      if (c=len) then
```

```
          maxi
    else
      let
          val hr1 = List.nth(q,maxi)
          val hr2 = List.nth(q,c)
          val cmp = compare(hr1,hr2,sm)
      in
          if (cmp = 1) then elect(q,maxi,c+1,sm)
          else elect(q,c,c+1,sm)
      end
end;
```

Functions *choose* receives a list of processes that are ready for running as first input parameter and scheduling method of the model as second parameter. It chooses the highest priority process from the list of input process by calling function *elect* and returns it index. Functions *choose* returns index of winning process for running as the result.

```
fun choose(q:ProcList, sm:SCHTYPE)=
let
    val mx = elect(q,0,1,sm)
in
    List.nth(q,mx)
end;
```

Function *remaining* receives list of processes that are ready for running as first input parameter and scheduling method of the model as second parameter. It calls function *elect* for taking index of the process that has highest priority for running. It creates a new list of processes that is same as the input list of processes except the one that had highest priority is removed from it as result. This function returns a list of input processes that do not selected for running.

```
fun remaining(q:ProcList, sm: SCHTYPE)=
let
    val mx = elect(q,0,1,sm)
in
    List.take(q,mx)^^List.drop(q,mx+1)
end;
```

Function *setExecutionProp* receives a process as input parameters. It calls the function *curTime* and records its value as start time field of input process and return updated input process as result. This function sets execution start time of the process that is selected to be run.

```
fun setExecutionProp(q:Process) : Process =  Process.set_ES q (curTime())
```

Function *isIdle* is the last function that is used in model of the system. Function *isIdle* indicates that whether the processor is in idle mode or not. This function takes three input parameters. First parameter is a list of processes that are ready for running. Second parameter is a list of the processes that will enter the system in future. The third parameter is a list of single process that is running now. We assumed that a single processor can run at most one process at a time. So, this list will contain at most one element at a time. If any process is running now or list of ready processes is not empty, then processor cannot be in idle state and function returns false as result. If no process is running and list of ready processes be empty and input time of no process in the list of input processes for running is reached based on the current simulation time, then function returns true as result.  If calling of function *exists* with input parameter of input process's list returns true, this means that based on the current simulation time, there exists at least one process that it's input time is reached. Therefore function *isIdle* will return false value.

```
fun isIdle(rq,init,run)=
if (run=[] andalso rq=[] andalso init<>[]) then
        if( not(exists(init))) then  true
        else false
 else
        false;
```

# 5. DETAILED OPERATIONS OF MODEL

This section of the paper explains transitions, guard condition of them and arcs inscriptions. Beyond that, detailed description of the model is presented. Basic model of the system is presented in figure 2. Location ReadyQueue in figure 2 keeps list of processes that their execution time are reached and are waiting for scheduling and running. Places Running1 and Running2 are fusion places and contain the same list of running processes. However, in the model there is only one processor and this list will contain at most one process.

Transition *Activate* will be enabled when place *NewTasks* contain the processes that their input time to the system is reached. This condition will be evaluated by calling function *exists*. By firing this transition, function *select* extracts some processes from list of processes in place *NewTasks* that their input time to system is reached. Then it adds them to end of the list of processes of place *ReadyQueue*. By firing of this transition, function *remove* eliminates some processes from list of place *NewTasks* that their input time to system is reached.

Transition *Dispatch* will be abled when, (1) no process is running (2) list ready processes is not empty (3) there is not exist any process in place *NewTasks* that their input time to system is reached, but is not entered to the list of ready processes. Reason of third condition is that when a process executes, the simulation time will advanced in amount of running time of that process. May be input time of a process is reached in this interval. To make sure that such processes are not omitted, if arrival input of a process is reached, transition *Dispatch* will be disabled and transition *Activate* will be enabled so that this process will be removed from list of processes that are waiting to input the system and will be added to the ready list. When guard condition of transitions *Dispatch* is being checked, priority of processes will be updated by function *update*.
By firing of the transition *Dispatch*, function *choose* selects the process with the highest priority and inserts it to list of place *Running1*. The function *remaining*, removes this selected process from list of processes of place *ReadyQueue* and holds rest of the list.

If a process is appeared in the list of place *Running1*, this means that this process is selected for dispatching and running. Transition *Execute* will be abled in this case. Firing of this transition represents running of this selected process. Simulation time will advanced in execution amount of this running process. Scheduling and running information of current executing process will be updated with function ***setExecutionPropoerty*** and will be added to the end of list of processes that their execution is finished and is hold in place *Finished*.

# 6. RESULTS OF SIMULATIONS

After running the simulation with input processes of Table 1 using SJF nonpreemptive scheduling, marking of place FinishedTasks in Figure 1 would be as follows.

1`[{PI=6,IT=1,ST=3,WT=0,ES=1,PR=(1,0)},{PI=4,IT=5,ST=2,WT=0,ES=5,PR=(5,0)},{PI=1,IT
=6,ST=4,WT=1,ES=7,PR=(6,0)},{PI=2,IT=7,ST=3,WT=4,ES=11,PR=(7,0)},{PI=3,IT=8,ST=2,
WT=6,ES=14,PR=(8,0)},{PI=5,IT=9,ST=3,WT=7,ES=16,PR=(9,0)}]@19

Figure 4 show Gantt chart of running processes using FCFS scheduling method. This figure is derived from the above marking.

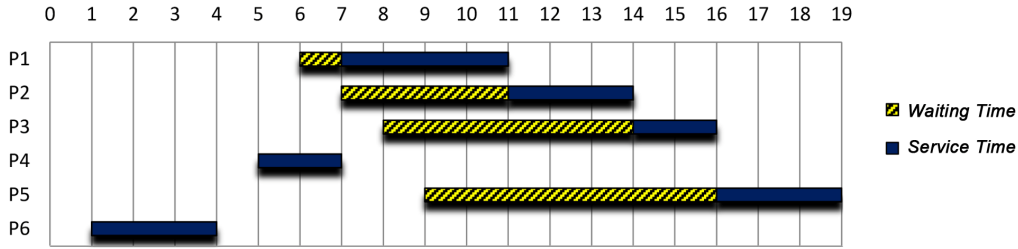

Figure 4.  Gant chart of FCFS scheduling of input processes.

After running the model with input processes of Table 1 using HRRN non-preemptive scheduling, marking of place FinishedTasks in Figure 1 would be as follows.

1`[{PI=6,IT=1,ST=3,WT=0,ES=1,PR=(100,0)},{PI=4,IT=5,ST=2,WT=0,ES=5,PR=(100,0)},{PI
=1,IT=6,ST=4,WT=1,ES=7,PR=(125,0)},{PI=3,IT=8,ST=2,WT=3,ES=11,PR=(250,0)},{PI=2,IT
=7,ST=3,WT=6,ES=13,PR=(300,0)},{PI=5,IT=9,ST=3,WT=7,ES=16,PR=(333,0)}]@19

Figure 5 shows Gantt chart of running processes using HRRN scheduling method.

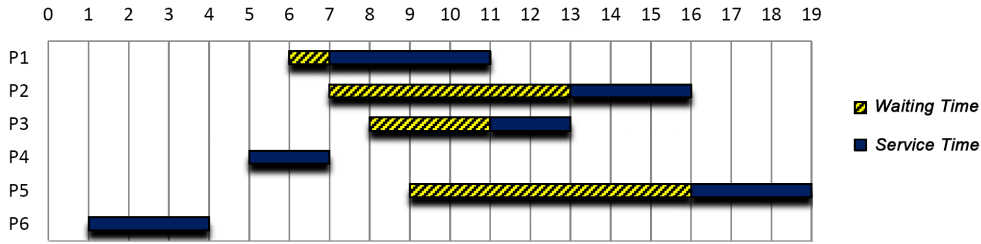

Figure 5.  Gant chart of HRRN scheduling of input processes.

Despite of most existing models that input tasks must be available at start time of system, No restriction on input time of tasks to system exits in current proposed model. Idle time of processor is considered in the model and can be calculated by analyzing final marking of place FinishedTasks. Waiting time and turnaround time of each task can be computed individually. This model can be used for verification all required properties of system. This model can be used in scheduling of real-time tasks. Model checking of system using formulas of Linear Temporal Logic (LTL) or Computational Tree Logic (CTL) permits us to prove that can system finishes a specific task before a predetermined deadline or not? Modular and hierarchical design of model permits easy modification and upgrading of the model in future works.

As next future work, we are intended to add external signal that enables or disables the processor. Modeling of single processor using preemptive scheduling method is under development too.

Proposed model is designed using Colored Petri net. Colored Petri net has features for simulation and performance evaluation of systems. But this modeling language is appropriate for system validation and sometimes is used for evaluation of systems. Therefore, performance evaluation of system using CPN tool is a little bit difficult in comparison with other specialized simulation tools. Model checking of system using CPN tools is a little difficult. Coding using predefined functions of modeling language and ML language is required for this purpose.

# 7. CONCLUSION

No unified approach for modeling enterprise architecture exists. Various aspects and different areas of the enterprise architecture can be modeled using various methods, tools and techniques. Different models of scheduling a single agent or processor using traditional or colored Petri net are presented in different papers. Most of presented models are simple and not parametric from this point that scheduling method is fixed in their models. In this paper a unified model is presented for a single processor that accepts a non-preemptive scheduling method as its input parameter. Model can accepts four famous non-preemptive scheduling methods in the form input token. Presented model can simulate scheduling of input tasks and also determines idle times of the processor.

Most researchers have used the classic Petri nets for modelling systems in single-level level of abstraction. However, hierarchical model of system in two levels of abstraction are presented in this paper. Hierarchical model of system using colored Petri net makes it possible to selectively paying attention to details of the system. Modular design is very useful way for dealing with complexities of for very large systems.

Petri net models of complex systems usually are very big. Most functionality of presented system are developed using coding of appropriate functions in the model. Using coding with ML language that colored Petri net benefits from it has many advantages that some of them are: 1) greatly decreases the size of the model, 2) reduces the size of state space of the system 3) decrease running time of the model, 4) decreases required memory for running of the model, 5) increase readability of the model, and 6) increase scalability of the model.

As future work, authors are intended to develop parameterized preemptive scheduling model of a single processor or agent. Enterprise architecture of most organizations are composed of a series of processors or agents and a series of workflows for tasks that flows from these processors. By hierarchical combining of different processors we can model wide range of enterprise architectures.

# REFERENCES

[1] Zachman, J. A. (1997) *Concepts of the framework for enterprise architecture*, *Zachman International, Inc., La Cañada, C*A.

[2] Van der Aalst, W. M., Stahl, C., & Westergaard, M. (2013) "Strategies for modeling complex processes using colored petri nets", in *Transactions on Petri Nets and Other Models of Concurrency VII*, ed: Springer, pp 6-55.

[3] Johnson, P., Johansson, E., Sommestad, T., & Ullberg, J. (2007) "A tool for enterprise architecture analysis", in *11th IEEE International Conference on Enterprise Distributed Object Computing 2007* (*EDOC 2007*), pp 142-142.

[4] Lankhorst, M. (2013) *Enterprise architecture at work: Modelling, communication and analysis*, Springer.

[5]     Van der Aalst, W. M. (1998) "The application of Petri nets to workflow management", *Journal of circuits, systems, and computers*, Vol. 8, No. 1, pp 21-66.

[6]     Van der Aalst, W. M. (1996) "Three good reasons for using a Petri-net-based workflow management system", in *Proceedings of the International Working Conference on Information and Process Integration in Enterprises (IPIC'96)*, pp 179-201.

[7]     Yilmaz, B. (2008) "Applications of Petri nets", M.Sc. Thesis, Izmir, Turkey.

[8]     Ha, S., & Suh, H.-W. (2008) "A timed colored Petri nets modeling for dynamic workflow in product development process", *Computers in industry,* Vol. 59, No. 2-3, pp 193-209.

[9]     Naedele, M. (1998) "Petri net models for single processor real-time scheduling", ed: Citeseer.

[10]   De Carvalho, M. F. H., & Haddad, R. B. B. (2012) "Production scheduling on practical problems", Production Scheduling on Practical Problems, Production Scheduling, Prof. Rodrigo Righi (Ed.), InTech, Available from: http://www.intechopen.com/books/production-scheduling/production-scheduling-on-practical-problems

[11]   Bożek, A. (2012) "Using timed coloured Petri nets for modelling, simulation and scheduling of production systems", *Production Scheduling,* Prof. Rodrigo Righi (Ed.), InTech, Available from: http://www.intechopen.com/books/production-scheduling/using-timed-coloured-petri-nets-for-modelling-simulation-and-scheduling-of-production-systems.

[12]   Zhang, Y., & Zhu, J. (2009) "Product line system modeling of the cold-rolled mill based on the hierarchy colored petri nets", in *IEEE International Conference on Automation and Logistics ( ICAL'09)*, pp 1553-1557.

[13]   Zhang, H., Gu, M., & Song, X. (2008) "Modeling and Analysis of Real-Life Job Shop Scheduling Problems by Petri nets", in *41st Annual Simulation Symposium (ANSS 2008)*, pp 279-285.

[14]   Zachman, J. A. (2003) *The Zachman framework: A primer for enterprise engineering and manufacturing* (electronic book), *www.zachmaninternational.com*.

[15]   Pinedo, M. (2012) *Scheduling: theory, algorithms, and systems*, Springer.

[16]   Valente, J., & Gonçalves, J. F. (2009) "A genetic algorithm approach for the single machine scheduling problem with linear earliness and quadratic tardiness penalties", *Computers & Operations Research*, Vol. 36, No. 10, pp 2707-2715.

**Authors**

**Saeid Pashazadeh** is Assistant Professor of Software Engineering and chair of Information Technology Department at Faculty of Electrical and Computer Engineering in University of Tabriz in Iran. He received his B.Sc. in Computer Engineering from Sharif Technical University of Iran in 1995. He obtained M.Sc. and Ph.D. in Computer Engineering from Iran University of Science and Technology in 1998 and 2010 respectively. He was Lecturer in Faculty of Electrical Engineering in Sahand University of Technology in Iran from 1999 until 2004. His main interests are modelling and formal verification of distributed systems, computer security, wireless sensor/actor networks, and applications of artificial neural networks. He is member of IEEE and senior member of IACSIT and member of editorial board of journal of electrical engineering at University of Tabriz in Iran. 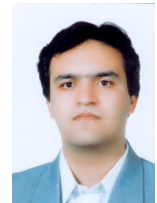

**Elham Abdolrahimi Niyari** is M.Sc. student of Information Technology Management in the field of Advanced Information Systems at University College of Mizan in Iran. She received her B.Sc. degree in Business Management from Payam Noor University of Osku in 2009. She's research interests are in the fields of information systems, business process management, simulation, Petri net, process modelling and workflow management systems. 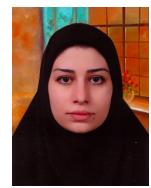